\documentclass[11pt,twoside]{article}
\usepackage{asp2010}

\usepackage{color}
\usepackage{soul}
\definecolor{jaune}{rgb}{1.0, 1.0, 0.0}

\resetcounters

\begin{document}

\title{Solving the Puzzle of the Massive Star System $\theta^2$\,Orionis\,A}
\author{V.~Petit,$^1$ M.~Gagn\'e,$^1$ D.~H.~Cohen$^2$, R.~H.~D.~Townsend$^3$, M.~A.~Leutenegger$^4$, M.~R.~Savoy$^1$, G.~Fehon$^1$ and C.~A.~Cartagena$^1$
\affil{$^1$Dept. of Geology \& Astronomy, West Chester University, USA}
\affil{$^2$Dept. of Physics \& Astronomy, Swarthmore College, USA}
\affil{$^3$Dept. of Astronomy, University of Wisconsin-Madison, USA}
\affil{$^4$Laboratory for High Energy Astrophysics, NASA/GSFC, USA}}

\begin{abstract}

The young O9.5~\,V spectroscopic binary $\theta^2$\,Ori\,A shows moderately hard X-ray emission and relatively narrow X-ray lines, suggesting that it may be a Magnetically Confined Wind Shock (MCWS) source, similar to its more massive analogue $\theta^1$\,Ori\,C. X-ray flares occurring near periastron led to the suggestion that the flares are produced via magnetic reconnection as magnetospheres on both components of the $\theta^2$\,Ori\,A binary interact at closest approach. 

We use a series of high-resolution spectropolarimetric observations of $\theta^2$\,Ori\,A to place an upper limit on the magnetic field strength of 135~G (95\% credible region). Such a weak dipole field would not produce magnetic confinement, or a large magnetosphere. A sub-pixel analysis of the {\it Chandra} ACIS images of $\theta^2$\,Ori\,A obtained during quiescence and flaring show that the hard, flaring X-rays are offset from the soft, quiescent emission by $0.4$\,arcsec. If the soft emission is associated with the A1/A2 spectroscopic binary, the offset and position angle of the hard, flaring source place it at the location of the intermediate-mass A3 companion, discovered via speckle interferometry.
The spectropolarimetric and X-ray results taken together point to the A3 companion, not the massive A1/A2 binary, as the source of hard, flaring X-ray emission.

We also discuss a similar analysis performed for the magnetic Bp star $\sigma$\,Ori\,E. We find a similar origin for its X-ray flaring. 

\end{abstract}

\section{Introduction}
The standard framework for understanding X-ray emission from effectively single O and early B stars is the Embedded Wind Shock (EWS) mechanism, in which instabilities in the radiation driving of the wind lead to shock-heating of a small fraction of the wind mass \citep{1988ApJ...335..914O,1997A&A...322..878F,1999ApJ...520..833O}. The associated X-ray emission is characterised by its relatively soft spectral energy distribution and by Doppler-broadened emission lines consistent with the wind velocity traced by UV resonance lines.  However, a small subset of O-type stars, with large-scale magnetic fields, generate X-rays via the Magnetically Confined Wind Shock (MCWS) mechanism \citep{1997ApJ...485L..29B,1997A&A...323..121B,2002ApJ...576..413U}. The prototype of this class, $\theta^1$\,Ori\,C (OV 7), has stronger and harder X-ray emission, and narrower X-ray emission lines, than EWS sources \citep{2005ApJ...628..986G}.

$\theta^2$\,Ori\,A is a hierarchical triple system, containing the second O-type star in the Orion Nebula Cluster.
It is a single-lined spectroscopic binary (component A1/A2) with an orbital period of 20\,d \citep{1991ApJ...367..155A}. 
The O9.5\,V primary (A1) has an estimated mass of 29\,$M_\odot$.
The A3 component is a visual companion detected through K-band speckle interferometry, with a flux ratio of 
$0.08\pm0.02$, a separation of $383\pm10$\,mas and a position angle of $291\pm2$\,deg \citep{1999NewA....4..531P}. The mass was estimated at 7\,$M_\odot$, although a solution with 3\,$M_\odot$ was also possible, and the star is probably still on the pre-main sequence.

The X-rays of $\theta^2$\,Ori\,A stand out from those of normal O-type stars.
It has moderately hard X-ray emission and  relatively narrow X-ray lines, and has thus been identified as an MCWS candidate, by analogy to $\theta^1$\,Ori\,C \citep{2011ApJ...734...14M}.  
Hard X-ray flares that may occur preferentially near periastron led \citet{2006ApJ...653..636S} to suggest an exotic new X-ray emission mechanism: binary-driven magnetic reconnection. 
It has also been proposed that X-ray flaring in magnetic massive stars could be caused by centrifugally-driven magnetic breakout, which is predicted to occur in the MCWS scenario when the magnetosphere mass reaches a critical level \citep{2006ApJ...640L.191U}.  
The possibility that the flaring could arise from the lower-mass visual companion was usually discarded based on the rarity of low-mass X-ray sources emitting at this level. However, it has been shown that there are some pre-main sequence stars in the Carina complex with X-ray emission as high as the bulk of single early OB stars \citep{2011ApJS..194....5G}

%In order to assess the applicability of the different X-ray production mechanisms, the magnetic field properties of $\theta^2$\,Ori\,A need to be determined. Furthermore, the effect of the binary companion(s) on the observed X-rays also needs to be evaluated.
In these paper, we assess the applicability of the different X-ray production mechanisms and we provide two new contributions to our understanding of the $\theta^2$\,Ori\,A system: (1) strong, new constraints on the magnetic field properties of the star, and (2) a sub-pixel spatial analysis of the X-ray emission from the system during both quiescence and flaring. 

%\citet{2002ApJ...574..258F} reported two ACIS observations from the \textit{Chandra} X-ray Observatory, separated by a few months. In the first observation, the count rate is nearly constant. However, the second observation had a count rate 4 times higher, followed by a \hlj{slow(rapid?)} decay over the 14\,h observation, typical of a flare. Combined with the hardening of the spectrum, this higher count rate translated in an increase of the X-ray luminosity by nearly an order of magnitude. 

\section{Magnetic observations}

We obtained observations with the ESPaDOnS spectropolarimeter located at the Canada-France-Hawaii Telescope in 2006 January, 2007 March and 2012 February. Our data consist of high resolution ($R\sim65\,000$), high signal-to-noise ratio spectra in intensity (Stokes $I$) and circular polarisation (Stokes $V$).  Each polarisation sequence consists of a set of four sub-exposures taken in different polarimeter configurations. Diagnostic null polarisation spectra ($N$) are calculated by combining the exposure in such a way that stellar polarisation cancels out, providing a check against spurious instrumental signal. 
For a complete description of the observing and reduction procedure with the \textsc{upena} pipeline using the \textsc{libre-esprit} software, see \citet{1997MNRAS.291..658D}. 

$\theta^2$\,Ori\,A is now a double-lined spectroscopic binary. The high S/N of our observations enable us to clearly distinguish the very broad ($\sim300$\,km\,s$^{-1}$) lines of the spectroscopic companion A2 in the He\,\textsc{i} lines (see Fig.~\ref{PetitV_ref:fig1}, left). From the radial velocities, we derive a mass ratio $M_1/M_2\sim1.8$, and a system mass $M_1+M_2\ga45$\,$M_\odot$, in the case of a circular orbit. 

\articlefiguretwo{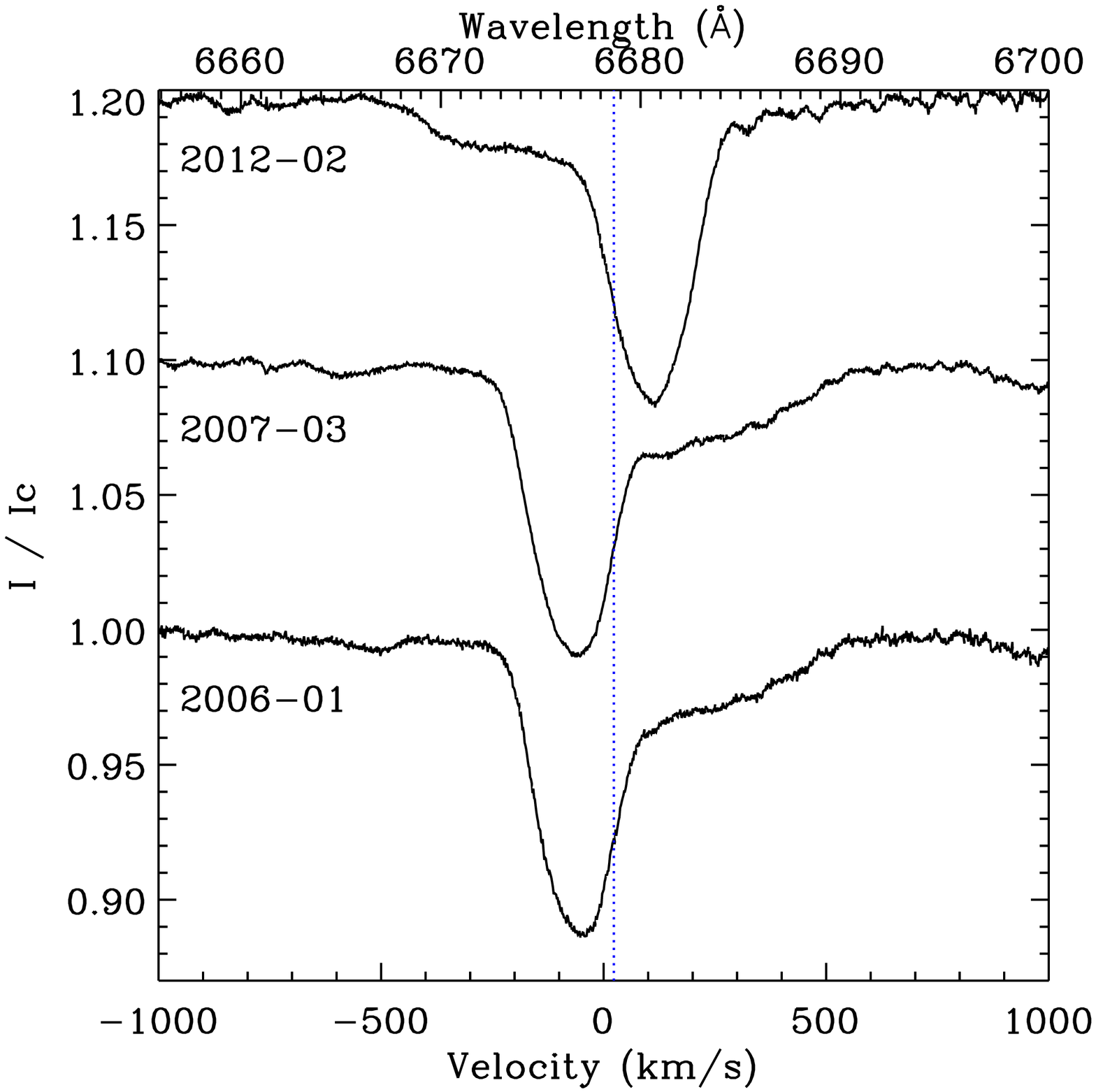}{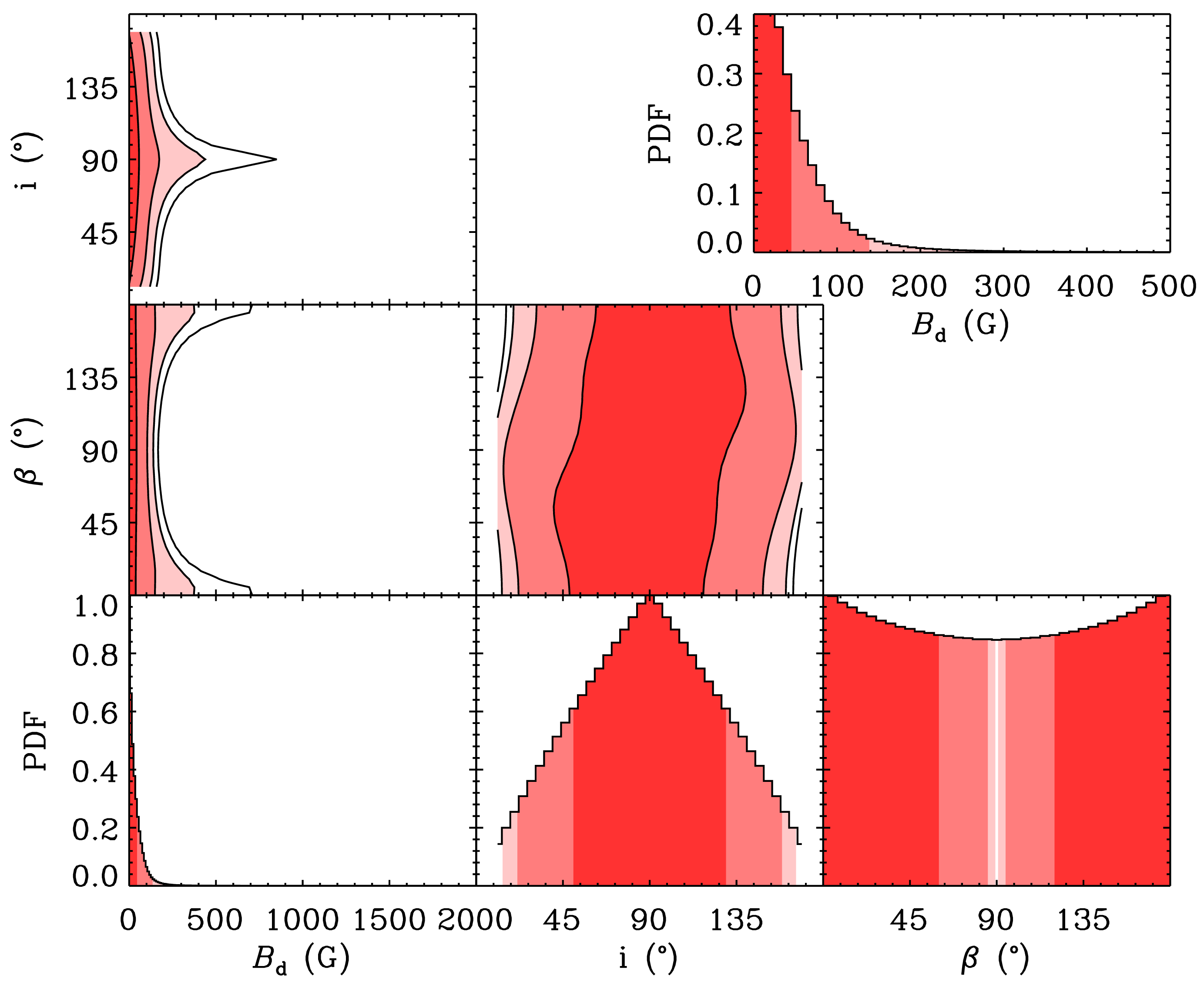}{PetitV_ref:fig1}{\textit{Left:} Stokes $I$ ESPaDOnS observations of $\theta^2$\,Ori\,A, showing the contribution of the spectroscopic companion A2. The dashed line correspond to the systemic velocity of the Orion Nebula Cluster. \textit{Right:} Posterior probability density functions (PDF) for the dipole model of $\theta^2$\,Ori\,A. The bottom row shows the marginalised PDFs for the dipole field strength, the rotation axis inclination and the magnetic obliquity from left to right, respectively. A zoomed-in version is shown in the upper right for the dipole strength PDF. The top left plots show the 2-D posteriori probability density marginalised for the $B_p$-$i$, $B_p$-$\beta$ and the $i$-$\beta$ planes. The 68.3\%, 95.4\%, 99.0\% and 99.7\% credible region are shaded in dark to pale colours respectively.}

The Least Squares Deconvolution technique \citep[LSD,][]{1997MNRAS.291..658D} was applied to our spectra, in order to extract a mean profile of higher S/N ratio. We use the iLSD code described by \citet{2010A&A...524A...5K}.
No circular polarisation signal was detected in our spectra. In order to assess the maximum possible field strength, we employed the method described by \citet{2012MNRAS.420..773P} which models the data assuming a centered dipolar magnetic field, using a Bayesian framework. Fig.~\ref{PetitV_ref:fig1} (right) shows the 1D and 2D posterior probability densities for the model parameters: the dipolar field strength $B_p$, the inclination of the rotational axis to the line of sight ($i$) and the obliquity of the magnetic axis to the rotational axis ($\beta$). The  upper limit to the 95.4\% credible region for $B_p$ is 135\,G, corresponding to the generally detectable field strength.

The degree of wind magnetic confinement $\eta_\star=B^2_\mathrm{eq}R^2/\dot{M}v_\infty$ describes the ratio of magnetic energy density and kinetic energy density of the wind \citep{2002ApJ...576..413U}. 
The stellar parameters of $\theta^2$\,Ori\,A were derived by \citet{2006A&A...448..351S} through NLTE modelling with \textsc{fastwind} ($T_\mathrm{eff}=35$\,kK, $\log g=4.1$, $\log(L/L_\odot)=4.93$, $R=8.2\,R_\odot$, $M=29\,M_\odot$). Using these parameters and the recipe of \citet{2000A&A...362..295V,2001A&A...369..574V} for the wind momentum ($\log\dot{M}=-7.1$, $v_\infty=2900$\,km\,s$^{-1}$), our dipole field ($B_\mathrm{eq}=\frac{1}{2}B_p$) limit leads to $\eta_\star\sim1$. Therefore, we do not expect any significant magnetic confinement for the O-type star primary of the $\theta^2$\,Ori\,A system.

%If we use our upper limit of 125G and assume of R, etc, etc 
%A field of 125\,G would not lead to significant magnetic confinement ($\eta_\star\sim1$).  

\section{X-ray observations}

The \textit{Chandra} X-ray Observatory's excellent spatial resolution (0.5\,arcsec) is unsurpassed among current and past higher-energy facilities. Although the ACIS detector (0.492\,arcsec pixels) under-samples the PSF of the high-resolution mirror assembly, event positions can be recovered with more accuracy than the pixel grid, based on physical models of the impact charge distribution among the CCD pixels \citep{2003ApJ...590..586L,2004ApJ...610.1204L}. These subpixel event repositioning techniques are now included in the standard ACIS observation reduction, and can be used to resolve structures at sub-pixel scales \citep[e.g.][]{2011ApJ...736...62W}.

Fig.~\ref{PetitV_ref:fig3} shows a colour image of the ACIS-I observation obtained as part of the \textit{Chandra} Orion Ultradeep Project \citep[COUP,][]{2005ApJS..160..319G}, a 838-ks exposure of a single field in the Orion Nebula Cluster. 
During these observations, the light curve of $\theta^2$\,Ori\,A showed low-level flaring \citep{2005ApJS..160..557S}. 
From this image, it is clear that the soft and hard emission are not co-spatial. For comparison, the relative positions of the A1/A2 spectroscopic binary and the A3 visual companions are indicated with white circles. 

\articlefigure{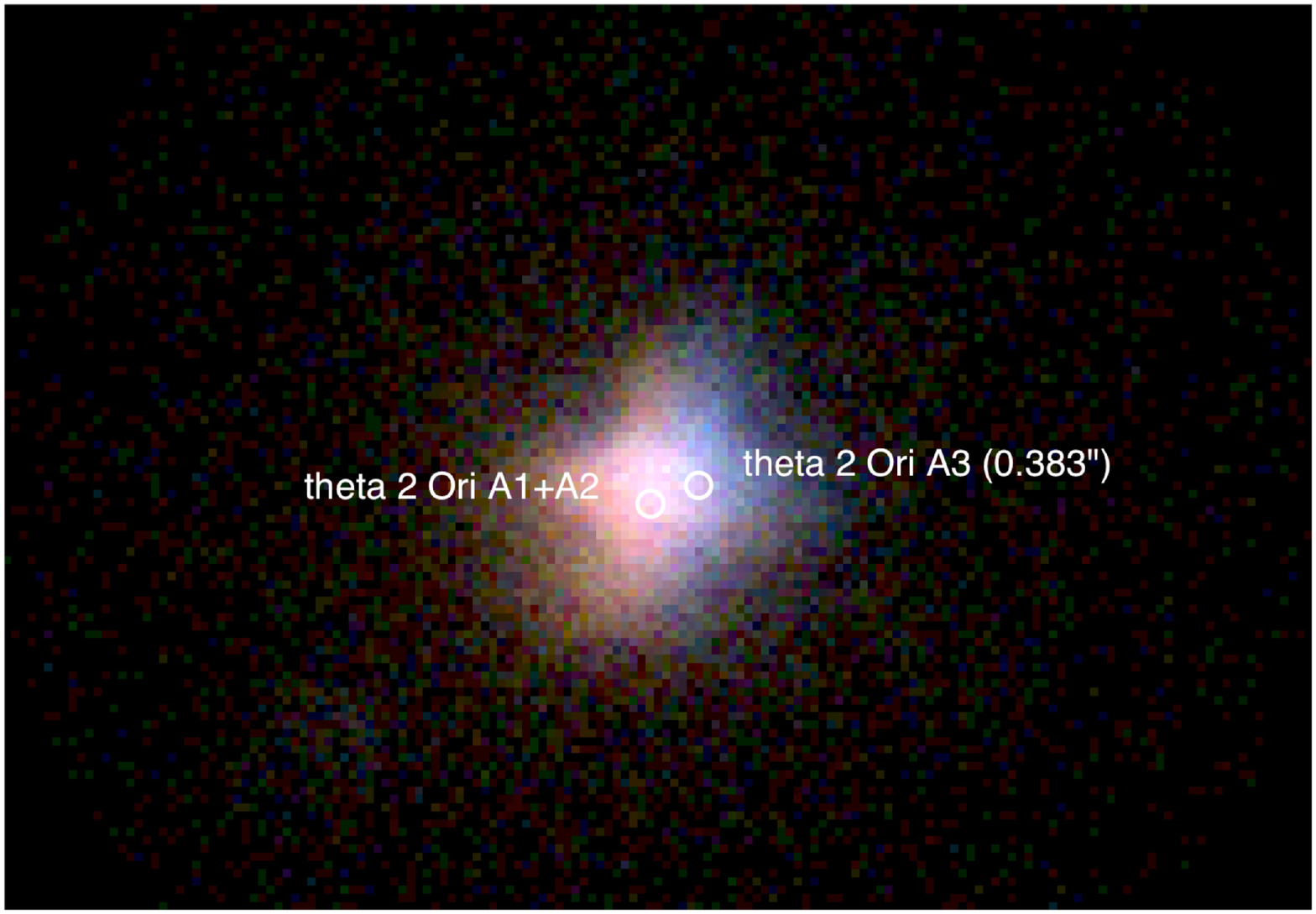}{PetitV_ref:fig3}{Colour image of the X-ray emission of $\theta^2$\,Ori\,A from the \textit{Chandra} Orion Ultradeep Project. The red, green and blue colours correspond to the soft (0.3-1.5\,keV), medium (1.5-2.5\,keV) and hard (2.5-8.0\,keV) bands, respectively. Note the offset between the soft and hard emission. The white circles indicate the relative position of the A1/A2 spectroscopic binary and the A3 visual companion. }

To improve the image quality, we performed a Richarson-Lucy deconvolution on two images containing the soft counts below 1.0\,keV and the hard counts above 1.2\,keV. We first used a synthetic PSF created from ChaRT rays fed into \textsc{marx} and \textsc{marxpileup}. Although this synthetic PSF provided good deconvolution at the pixel level, we found that it is not accurate enough at the subpixel level and was leading to deconvolution artefacts. We therefore used a fainter nearby source (COUP\,1143) as a PSF template. The deconvolution, shown in Fig.~\ref{PetitV_ref:fig4} (left), is not perfect because the PSF template and the source do not have the same pile-up, which affects mostly the core of $\theta^2$\,Ori\,A. Nonetheless a clear separation between the soft and the hard components is found. 

\articlefiguretwo{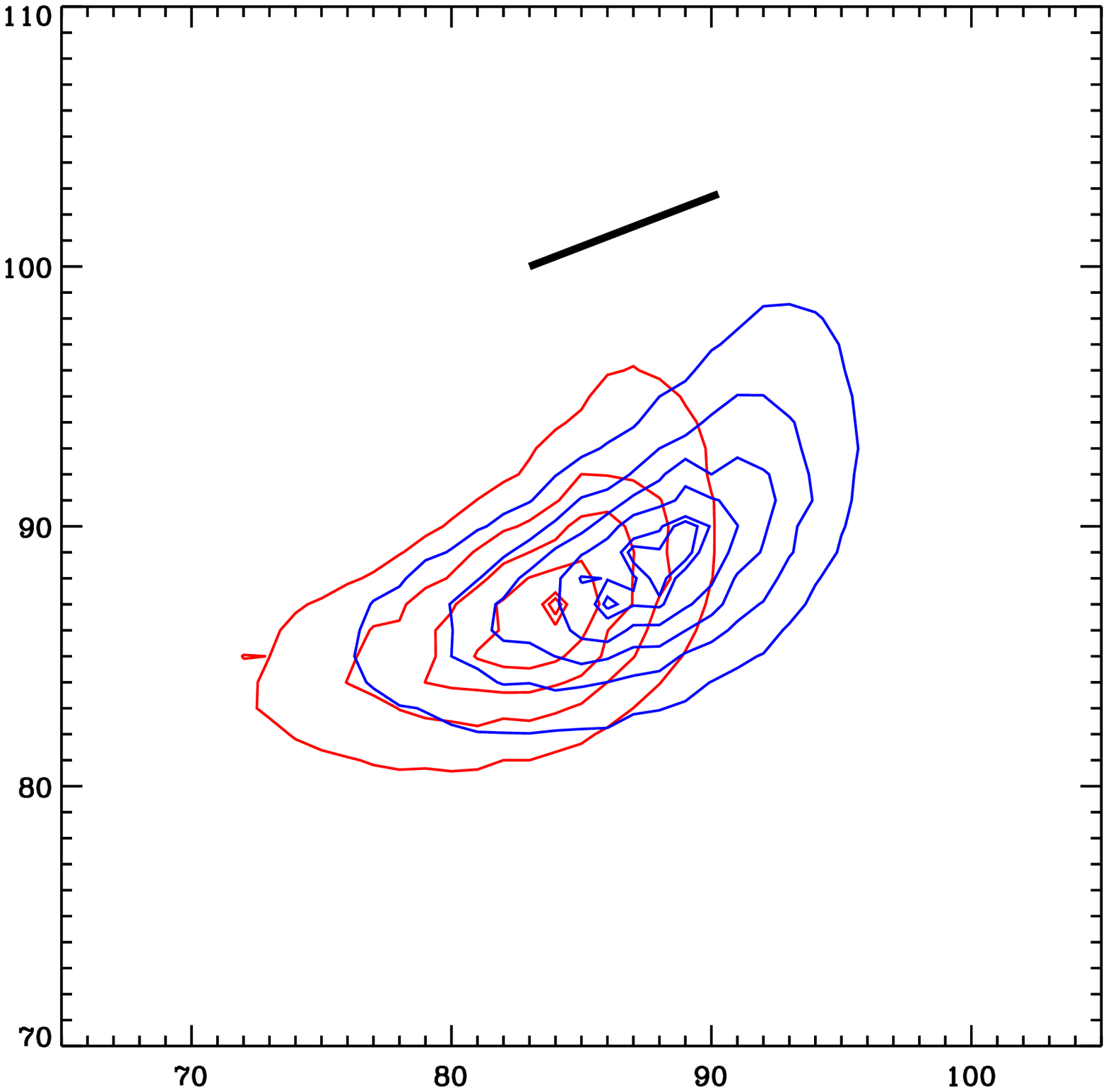}{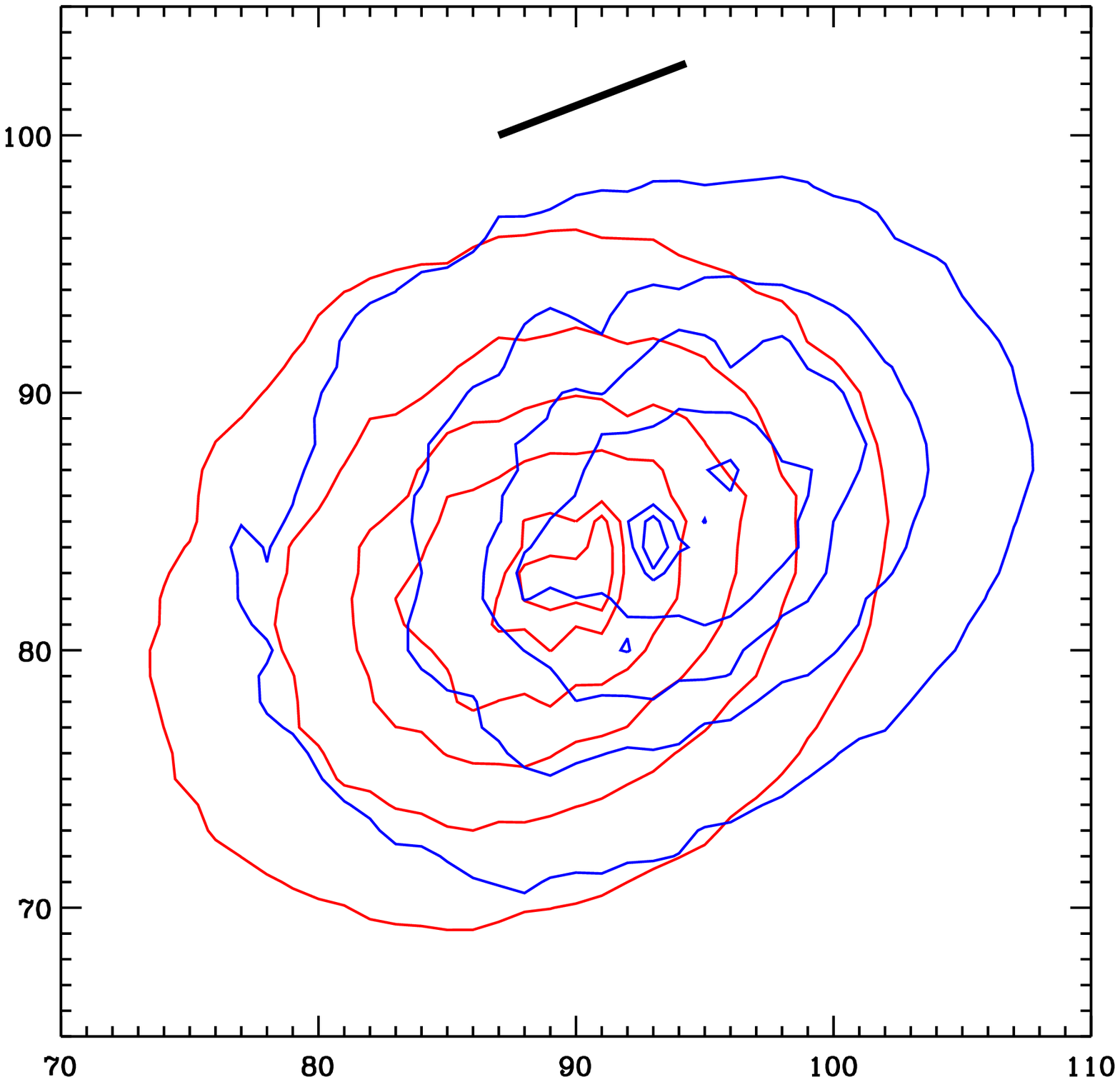}{PetitV_ref:fig4}{Richardson-Lucy deconvolution of the ACIS-I COUP observations (left) and the zeroth order ACIS-HETG observations (right). The red and blue contours represent the soft band (0.3-1.0\,keV) and the hard band (1.2-7.0\,keV), respectively. The black lines represent the separation and position angle of the A1+A2 spectroscopic binary and the A3 visual companion. The axes are in unit of 0.05\,arcsec ACIS sub-pixels.}

%In Fig.~\ref{PetitV_ref:fig4} (left), we show a Richarson-Lucy deconvolution, using a fainter near-by source (COUP\,1143) as a PSF template. 
%The deconvolution is not perfect because the PSF template and the source do not have the same pile-up. Nonetheless a clear separation between the soft and the hard components is found. 

%Although the deconvolution is not perfect, given that the PSF template and the source do not have the same pile-up, a separation is seen between the soft and the hard components. 

In addition to the COUP observations, $\theta^2$\,Ori\,A was observed multiple times with the HETG \citep[for a complete list see][]{2006ApJ...653..636S,2011ApJ...734...14M}. 
We selected all the available HETG observations with an off-axis angle less than 2.5\,arcsec, for a total of 347\,ks during which $\theta^2$\,Ori\,A was in quiescence. 
Approximatively half the counts in the HETG are undispersed and detected in the zeroth order image of the source on the ACIS-S chip. Because the source was not flaring, and because the effective area in the zeroth order is lower than the bare ACIS-I, pile-up was minimal in these ACIS-S data.
However, given with the shorter exposure time, the data have less S/N than the COUP observations. We therefore performed an adaptive smoothing on the PSF template source. 
The separation of the soft and hard emission is also seen in our deconvolution (Fig.~\ref{PetitV_ref:fig4}, right).

%We also perform the same analysis on the sum of the available zeroth order HETG observation during quiescence. We selected only observations with an off-axis angle less than 2.5\,arcsec, for a total of 300\,ks. The separation of the soft and hard emission is also seen in our deconvolution (Fig.~\ref{PetitV_ref:fig4}, right). 

Therefore, it seems that the A3 companion is contributing significantly to the hard X-ray emission of the $\theta^2$\,Ori\,A system, both during quiescence and during flaring episodes. Although there are some HETG and ACIS observations taken during two major flaring events \citep{2002ApJ...574..258F,2006ApJ...653..636S}, the short duration of these observations does not allow us to perform a deconvolution, at least until the sub-pixel PSF of \textit{Chandra} has been better characterised, to allow the use of synthetic PSFs. 

\section{$\theta^2$\,Ori\,A3}

Based on series of K-band speckle interferograms, \citet{1999NewA....4..531P} found 8 new companions around 7 primaries, from a target sample of 13 O and early-B stars in the Orion Nebula Cluster. For the companion of $\theta^2$\,Ori\,A (= Par\,1993), they estimate a mass of $7 M_{\odot}$ if A3 is close to the main sequence, and $3 M_{\odot}$ if it is a 0.3-Myr old pre--main-sequence (PMS) star. If  $\theta^2$\,Ori\,A is as old as 1\,Myr, then the A3 companion would be a $\sim 4\,M_{\odot}$ PMS star. Because of the age and extinction uncertainties, and the degeneracy in the mass estimates from the HR diagram, A3's luminosity is in the range 30-2\,000\,$L_{\odot}$, comparable to many Herbig Ae/Be stars.

\citet{2002ApJ...574..258F} argued that the giant flare seen on $\theta^2$\,Ori\,A with \textit{Chandra} in 2000 was probably not produced by a lower-mass companion. 
Since that time however, strong X-ray emission has been observed on a number of intermediate-mass Herbig Ae/Be stars: HR\,5999 (A7e), HR\,6000 (A1.5III) and Z\,CMa (F5e) \citep{1994A&A...292..152Z}, V380\,Ori \citep[A1e,][]{2006A&A...457..223S}, AB Aur \citep[B9.5e,][]{2007A&A...468..443T}, and SS73\,24=Hen\,3-485 \citep[Be pec,][]{2011ApJS..194....5G}.
Another example is M16 ES-1, a highly absorbed 200\,$L_\odot$, 4-5\,$M_\odot$ YSO at the head of Pillar 1 in the Eagle Nebula. Though it was not flaring in one 60-ks Chandra observation, its $L_\mathrm{X}$ is $10^{32}$\,erg\,s$^{-1}$ and $kT = 2.2\pm1.0$~keV \citep{2007ApJ...654..347L}.
% HD~242926=MCW~297 (O7~V) (Hamaguchi et al. 2000, Vink et al. 2005)
Strong, hard and sometimes flaring X-rays are seen for more-massive T-Tauri stars:
e.g., SU Aur \citep{2007A&A...471..951F}, V773\,Tau \citep{1998ApJ...503..894T},
DoAr\,21  \citep{2002ApJ...572..300I,2009ApJ...703..252J,2004ApJ...613..393G} and a handful of
intermediate-mass T-Tauri stars in the Orion Nebula Cluster \citep{2005ApJS..160..469F,2008ApJ...688..418G}.

Though the origin of strong X-ray emission on intermediate-mass stars is not fully understood \citep[see][for a thorough discussion]{2007A&A...468..443T}, the X-ray luminosity and flares from $\theta^2$\,Ori\,A3 appear be similar to the largest flares seen on intermediate-mass T-Tauri stars.

\section{The magnetic star $\sigma$\,Ori\,E}

These results raise interesting questions about the origin of X-ray flares reported on other massive stars, for example the large flare reported in \textit{XMM-Newton} observations of the magnetic Bp star $\sigma$\,Ori\,E \citep{2004A&A...421..715S} and proposed to be caused by centrifugal mass ejection and field reconnection \citep{2006ApJ...640L.191U}. $\sigma$\,Ori\,E was observed by \textit{Chandra}, with the HRC detector in 2002. Although the HRC has no usable spectral resolution, it has the advantage of not being affected by pile-up. Moreover, the bright nearby X-ray source $\sigma$\,Ori\,AB can be used as a high S/N PSF template. During this 98-ks exposure, $\sigma$\,Ori\,E shows an increase in count rate by a factor of two, during which the centroid of the source moves to the North-West. 

In addition to image deconvolution, we have developed an algorithm to directly test the possibility that the X-ray emission from $\sigma$\,Ori\,E is a blend of multiple sources. We model the HRC image as a superposition of $m$ sources ($m = 1,2,3$) each characterised by a position and an amplitude (Townsend et al, in prep). 
%We assume that all sources share the same PSF, consisting of an axisymmetric spline fit to the HRC image of $\sigma$\,Ori\,AB out to a radial distance of 2\,arcsec from the centroid, and an $r^{-2}$ curve for the wings at $>2$\,arcsec. 
For each choice of $m$, the position and strength of the sources are
determined by maximising the Bayesian posterior probability of the HRC
event list for $\sigma$\,Ori\,AB, subject to uniform prior probabilities
and likelihood functions based on the adopted PSF.

To assess whether the $m=1$, $2$ or $3$ models provide (relatively) 
better fits to the observations, we take ratios of Bayes factors (the
likelihoods marginalised over all possible parameter choices). The
$m=1$ (single-source) model is strongly disfavoured compared to the
$m=2$ and $m=3$ models. The $m=3$ model is more favoured than the $m=2$
case, but the third source accounts for only 8\% of the total X-ray
flux, and therefore appears to be an artefact of over-fitting the
data (not located at the same position angle than the PSF hook\footnote{http://cxc.harvard.edu/ciao4.4/caveats/psf\_artifact.html}). 
This leaves the $m=2$ model, for which the sources have a position angle of 
$305^{\circ}$ (North via East), a separation of $0.42$\,arcsec and an amplitude ratio $A_1/A_2=0.4$. 
Using
multi-conjugate adaptive optics in the $H$ and $K_{\rm s}$ bands,
optics imager, \citet{2009A&A...493..931B} report a low-mass ($M \la
1\,M_{\odot}$) companion to $\sigma$\,Ori\,E at a position angle of
$\sim 300^{\circ}$ and a separation of $\sim 0.33$\,arcsec. Given the level
of agreement between these values, we identify the second source with the
companion found by \citeauthor{2009A&A...493..931B}

%\[
%\alpha_{1} = 5^\mathrm{h}38^\mathrm{m}47\fs194, \qquad
%\delta_{1} = -2^{\circ}35'40\farcs53, \qquad
%A_{1} = 0.29
%\]
%and
%\[
%\alpha_{2} = 5^\mathrm{h}38^\mathrm{m}47\fs171 \qquad
%\delta_{2} = -2^{\circ}35'40\farcs28, \qquad
%A_{2} = 0.71
%\]
%(here, celestial coordinates are J2000.0, and $A_{1,2}$ are the source strengths expressed as a fraction of the total X-ray flux). 
%For comparison, the \textit{PPM Catalog} coordinates of $\sigma$\,Ori\,E (which was \textit{not} observed by \textit{Hipparcos}) are $\alpha =5^\mathrm{h}38^\mathrm{m}47\fs194$, $\delta =-2^{\circ}35'40\farcs54$. These values are almost identical to those determined for source 1, and we therefore identify source 1 as $\sigma$\,Ori\,E itself.

\section{Conclusion}

The early-type massive system $\theta^2$\,Ori\,A has remarkable X-ray properties, including a hard flaring component that had been attributed to magnetically confined wind shocks, binary-driven magnetic reconnection or centrifugally-driven magnetic breakout. 

High S/N, high resolution spectropolarimetric measurements show no evidence of a magnetic field on $\theta^2$\,Ori\,A. The upper limit on the dipolar field strength of 135\,G (95\% credible region) would not lead to significant magnetic wind confinement for the O-type star primary. 
We have also analysed two datasets from the \textit{Chandra} X-ray Observatory. The first consists of a deep exposure of the Orion Nebular Cluster obtained for the \textit{Chandra} Orion Ultradeep Project. During the exposure, $\theta^2$\,Ori\,A showed low level flaring. The second dataset regroups all the HETG observations taken during quiescence with an off-axis angle less than 2.5\,arcsec. Leveraging the sub-pixel capability of the ACIS detector, we have shown that a significant part of the hard X-ray emission is not co-spacial with the soft emission, the separation and position angle corresponding to the position of the A3 visual companion. 
We therefore suggest that the hard flaring originates from the pre-main sequence visual companion, whereas the soft component originates primarily from the O-type star's wind. 
%This scenario was usually discarded based on the rarity of low-mass X-ray sources emitting at this level. However, it has been shown that there are some pre-main sequence stars in the Carina complex with X-ray emission as high as the bulk of single early OB stars \citep{2011ApJS..194....5G}.

These results raise interesting questions about the origin of X-ray flares reported on other massive stars, such as the Be star $\lambda$\,Eri \citep{1993ApJ...409L..49S}, the early B-type star $\theta^1$\,Ori\,A \citep{2005ApJS..160..557S} and the magnetic Bp star $\sigma$\,Ori\,E \citep{2004A&A...421..715S}, and whether massive stars can intrinsically flare.

%\acknowledgements VP acknowledges support from the fonds qu\'eb\'ecois de la recherche sur la nature et les technologies.
\section*{Discussion}

\textsc{Wade}: Rotational modulation of X-rays may be the only unambiguous way to ascribe X-rays to the hot star and rule out a companion as the source.

\noindent\textsc{Petit}: Rotational modulation of the X-rays, like the disk occultation of the X-ray emitting material trapped in magnetic closed loops of $\theta^1$\,Ori\,C, is indeed a strong confirmation of the MCWS mechanism. 
The ambiguity is related to sporadic flaring events, like those that could be expected from centrifugal mass ejection predicted by MCWS.

\bibliographystyle{asp2010}
\bibliography{PetitV}

\end{document}